\documentclass[a4paper]{article}
\usepackage{Odyssey2018}
\usepackage{epsfig,amssymb,amsmath}
\usepackage{url}
\ninept

\usepackage[font=small,skip=0pt]{caption}

\title{The Voice Conversion Challenge 2018:\\ Promoting Development of Parallel and Nonparallel Methods}

\makeatletter
\def\name#1{\gdef\@name{#1\\}}
\makeatother
\name{{\em Jaime Lorenzo-Trueba$^1$, Junichi Yamagishi$^{1,2}$, Tomoki Toda$^3$,}\\
      {\em Daisuke Saito$^4$, Fernando Villavicencio$^5$, Tomi Kinnunen$^6$, Zhenhua Ling$^7$}}

\address{$^1$ National Institute of Informatics, Japan 
$^2$ University of Edinburgh, UK 
$^3$ Nagoya University, Japan \\ 
$^4$ University of Tokyo, Japan 
$^5$ ObEN, USA 
$^6$ University of Eastern Finland, Finland \\
$^7$ University of Science and Technology of China, China \\ 
{\small \tt vcc2018@vc-challenge.org} }
\begin{document}
\maketitle
\begin{abstract} 

We present the Voice Conversion Challenge 2018, designed as a follow up to the 2016 edition with the aim of providing a common framework for evaluating and comparing different state-of-the-art voice conversion (VC) systems. The objective of the challenge was to perform speaker conversion (i.e.\ transform the vocal identity) of a source speaker to a target speaker while maintaining linguistic information. As an update to the previous challenge, we considered both parallel and non-parallel data to form the Hub and Spoke tasks, respectively. A total of 23 teams from around the world submitted their systems, 11 of them additionally participated in the optional Spoke task. A large-scale crowdsourced perceptual evaluation was then carried out to rate the submitted converted speech in terms of naturalness and similarity to the target speaker identity. In this paper, we present a brief summary of the state-of-the-art techniques for VC, followed by a detailed explanation of the challenge tasks and the results that were obtained.

\end{abstract}

\section{Introduction} 

Voice conversion (VC) is a technique to transform the speaker identity included in a source speech waveform into a different one while preserving linguistic information of the source speech waveform. VC has great potential in the development of various new applications such as a speaking aid for individuals with vocal impairments such as dysarthric patients \cite{kain07ds}, a voice changer to generate various types of expressive speech \cite{turk10ex}, novel vocal effects of singing voices \cite{villavicencio05svc}, silent speech interfaces \cite{toda12nam}, and accent conversion for computer assisted language learning \cite{felps09call}. 

VC is also known as an advanced presentation attack method used for automatic speaker verification systems. Many presentation attack detection methods have been proposed through the recent automatic speaker verification spoofing and countermeasures (ASVspoof) challenges \cite{7858696} to automatically discriminate such converted speech from genuine speech and to achieve secure and reliable speaker verification systems. The ASVspoof 2015 database \cite{7858696}, which includes various VC and speech synthesis attacks, is a based on the Spoofing and Anti-spoofing (SAS) corpus several authors of this paper had co-developed in 2014 \cite{7400997}. 

We launched the Voice Conversion Challenge (VCC) 2016 \cite{Toda+2016,Wester+2016} at Interspeech 2016. The objective of the challenge was to better understand different VC techniques built on a freely-available common dataset to look at a common goal and to share views about unsolved problems and challenges faced by current VC techniques. The VCC 2016 focused on the most basic VC task, that is, the construction of VC models that automatically transform the voice identity of a source speaker into that of a target speaker using a clean parallel training database where source and target speakers read out the same set of utterances in a professional recording studio. 17 research groups participated in the 2016 challenge. The challenge was successful and it established new standard evaluation methodologies and protocols for bench-marking the performance of VC systems. 

In 2018, we launched the second edition of the VCC, the VCC 2018. In this second edition, we revised three aspects of the challenge. First, we reduced the amount of speech data used for the construction of participants' VC systems to half. This is based on feedback from participants in the previous challenge and is also essential for practical applications. Second, we introduced a more challenging task, referred to as the Spoke task, in addition to a variation of the main task in the first edition, referred to as the Hub task. In the Spoke task, participants need to build their VC systems using a non-parallel database in which source and target speakers read out different sets of utterances. We then evaluate both parallel and non-parallel voice conversion systems via the same large-scale crowdsourcing listening test. Third, we attempted to bridge the gap between the automatic speaker verification (ASV) and VC communities. Since new VC systems developed for the VCC 2018 may be strong candidates for enhancing the ASVspoof 2015 database, we also assessed spoofing performance of the systems on the basis of anti-spoofing scores. In this paper, we describe the overview of the challenge and listening test results. The spoofing performance is described in \cite{tomivcc2analysis}. 

This paper describes the set-up of the VCC 2018. After briefly summarizing a basic VC framework needed for speaker conversion in Sect.\ 2, we will explain the tasks of the challenge, including the guidelines for participants, the details of the common dataset, and how the evaluation of diverse VC systems was designed in Sect.\ 3. The main results of the challenge are presented in Sect.\ 4. We conclude our research in Sect.\ 5. 

\section{Voice Conversion} 

\subsection{Basic Framework for Speaker Conversion}
To achieve speaker conversion, a data-driven approach is usually used to develop a conversion function to modify speech features including both segmental and prosodic features. Two main VC frameworks have been studied on the basis of the data-driven approach, parallel VC and non-parallel VC.

Parallel VC is the most standard framework. A parallel speech dataset consisting of utterance pairs of source and target speakers is used to develop the conversion function \cite{abe90vc}. A training dataset is usually developed by performing time frame alignment between the source and target voices in each utterance pair with dynamic time warping so that each time aligned frame pair shares the same linguistic information. These time aligned frame pairs are used as a supervised training dataset to develop the conversion function.

Non-parallel VC is a more valuable but challenging framework. This framework uses a non-parallel speech dataset in which the target speaker's utterances are different from those of the source speaker. Although no dominant method has been proposed yet, there are three typical approaches; 1) construction of a pseudo parallel dataset from the non-parallel speech dataset by performing acoustic clustering of the source and target speakers' data \cite{Suendermann2006}, 2) an adaptive approach that transforms a canonical conversion function estimated using existing parallel datasets into the source or target speakers separately using their utterances \cite{Mouchtaris2006}, and 3) an approach that estimates latent variables corresponding to speaker-independent phonemic content \cite{Hsu2017}.

\subsection{Feature Extraction and Conversion}
\textbf{Acoustic-to-acoustic mapping:} A typical VC framework uses high-quality speech analysis/synthesis techniques, such as a harmonic plus noise model (HNM) \cite{stylianou01hnm}, STRAIGHT \cite{kawahara99straight}, and WORLD \cite{Morise2016}. The speech features extracted from the source speaker's voice are converted to those of the target speaker directly, and then the converted speech features are used to generate a converted speech waveform. Therefore, it is necessary to carefully design conversion functions that transform only speaker identity.

\noindent\textbf{Phonetic posteriors-to-acoustic mapping:} It is also possible to use speaker-independent features capturing only linguistic information as another input in conversion such as phonetic posteriors estimated using a deep neural network (DNN)-based phone classifier \cite{Sun2016}. In this case, it is necessary to convert the linguistic features into the target acoustic features to generate the converted speech waveform. Since this conversion function can be constructed using only a speech dataset of the target speaker, this framework can be used for both the parallel and non-parallel VCs.

\noindent\textbf{Conversion function:} It is essential to use nonlinear regression functions for the above mappings. Many methods have been proposed: 1) a piecewise linear mapping (e.g., Gaussian mixture models (GMM) \cite{stylianou98gmm} and restricted Boltzmann machines (RBM) \cite{nakashika13rbm}), 2) nonlinear functions (e.g., kernel functions \cite{helander12dkplsr} and DNNs \cite{chen14dnn, SAITO20172017EDL8034}), and 3) an exemplar-based mapping (e.g., non-negative matrix factorization \cite{takashima13nmf, wu14nmf}). Moreover, various time-series methods have also been proposed to consider dynamic properties of the speech features \cite{toda2007voice,pilkington11gpr, xu14gpr,sun15blstm}. For more details, please refer to \cite{MOHAMMADI201765}.

\subsection{Waveform Generation}
\textbf{Deterministic vocoder:} A high-quality vocoder based on the source filter model is often used to generate a converted speech waveform from the converted speech features. However, quality degradation is often caused in the converted speech waveform due to approximations required in the source filter model. To alleviate this issue, various approaches have been proposed, such as improvements to an excitation model \cite{kain01phase} and an implementation of waveform shape modeling \cite{ye2006phase} and phase reconstruction called Griffin-Lim \cite{ref:Griffin84}. A vocoder-free approach has also been proposed, where an input speech waveform is directly filtered using a time variant filter \cite{Kobayashi2016}.

\noindent\textbf{Data-driven vocoder:} Recent progress in deep learning has made it possible to directly model speech waveform samples using neural networks, e.g., WaveNet \cite{Oord2016} and SampleRNN \cite{Mehri2016}. Inspired by these methods, a neural vocoder has been developed and its effectiveness has been confirmed \cite{tamamori2017speaker}.

\section{The Voice Conversion Challenge 2018}

\subsection{Hub Task} 

The objective of the challenge was speaker identity conversion. The dataset for the Hub task consisted of parallel corpora (same utterances) of four different sets of source and target speakers. The participants were asked to develop conversion systems and to produce converted data for all possible source-target pair combinations. Unlike the previous challenge, phonetic transcriptions were also included in the dataset, and the participants were allowed to use the these to train their conversion systems. A detailed description of the dataset is provided in the following section.  

The main guidelines to participate with an entry were as follows:
\begin{itemize}
\setlength{\itemsep}{0mm}
    \item Manual editing or system tuning in the conversion step was not allowed. Manual optimization of individual conversion systems was allowed only in the training stage. 
    \item Manual transcriptions (phoneme or linguistic information) of the evaluation data were not allowed. However, automatic speech recognition systems may be used to generate this information.
    \item The use of other source and target speaker's data included in the VCC 2018 dataset was allowed to develop a conversion system for a specific source-target pair.
    \item The transformation of any acoustic features, including supra-segmental and duration features, was allowed.
    \item The use of data other than the VCC 2018 dataset for training purposes was allowed.
    \item Participants were free to discard content (utterances) of the training set at their convenience.
    \item Participants were not allowed to submit multiple entries.
\end{itemize}

The Hub task was mandatory for all the participants. They were asked to submit their entry (only waveforms) after generating the converted materials from the evaluation data and to fill in a questionnaire to provide information and a description of their conversion system and their main related techniques to the organizers. Furthermore, the entries were evaluated in terms of target speaker similarity and naturalness using listening tests carried out by the organizers, as described in Section \ref{subsec:eval}. 

\subsection{Spoke Task} 
In the VCC 2018, non-parallel voice conversion was also evaluated. The dataset for the Spoke task has exactly the same target speaker's data. However, the source speakers were different from those of the Hub task and their utterances were also all different from those of the target speakers. The same guidelines as the Hub task were used in the Spoke task.

\subsection{Dataset} 

Like in the VCC 2016, the dataset used in the VCC 2018 is based on the device and production speech (DAPS) dataset \cite{DAPS}, which includes recordings of professional US English speakers in a professional studio without significant noise effects and is available online for free\footnote{\url{https://archive.org/details/daps_dataset}.}. The ``clean'' version of the original recordings, in which most of the non-speech sounds were removed manually, was used as the dataset in this challenge. The recorded audio includes about 13 minutes of speech sounds uttered by each of the 20 speakers. The recordings were down-sampled to 22.05 kHz for this challenge.

Source and target speakers for this challenge were different from those selected for the VCC 2016. 12 speakers, 6 female and 6 male speakers, were selected from the original pool of 20 in the DAPS dataset. Among the 12 speakers, we selected 2 female and 2 male speakers as target speakers common to the Hub and Spoke tasks. As done at the VCC 2016, one of the most important criterion for the selection of the speakers was to avoid those observing strong excitation (e.g. voice-quality) and/or prosody dependent perceptual cues in order to reduce their influence on the evaluation of the timbre similarity in the perceptual tests.

From the remaining speakers, we selected 2 female and 2 male speakers as source speakers in the Hub task and the final 2 female and 2 male speakers as source speakers in the Spoke task. The speakers to be used as target ones were decided subjectively by selecting the voices with the most distinctive timbre after perceptual inspection. We proceeded to use the speakers originally labelled on the DAPS set as (m5, m7, f1, f6) and (m2, m3, f2, f5) as the sets of source speakers in the Hub and Spoke tasks, respectively. They are denoted as (VCC2SM1, VCC2SM2, VCC2SF1, VCC2SF2) and (VCC2SM3, VCC2SM4, VCC2SF3, VCC2SF4) respectively for this challenge. We then used speakers originally labelled on the DAPS set as (m1, m8, f8, f10) as target speakers in both tasks. They are denoted as (VCC2TM1, VCC2TM2, VCC2TF1, VCC2TF2). 

Each of the source and target speakers have a set of 81 sentences. Again in the Hub task, the target and source speakers have the same set of sentences whereas the target and source speakers have a different set of sentences in the Spoke task. The number of test sentences for evaluation was 35 and the sentences were released to participants about one week before they were required to submit their converted voices. The participants were asked to build systems for all the 4$\times$4=16 combinations of source-target pairs in each of the Hub and Spoke tasks.
\subsection{Evaluation methodology}\label{subsec:eval}

Subjective listening tests were designed to perceptually evaluate the naturalness and speaker similarity of the converted samples for all speaker pairs considered in both the Hub and Spoke tasks. 

\textit{Naturalness}. Subjects were asked to evaluate the naturalness of voice converted samples and natural speech on a scale from 1 (completely unnatural) to 5 (completely natural).
    
\textit{Similarity}. To measure the similarity of VC samples, the Same/Different paradigm from the VCC 2016 \cite{Wester+2016} was used. Subjects were given two samples and asked the following: \emph{ ``Do you think these two samples could have been produced by the same speaker? Some of the samples may sound somewhat degraded/distorted. Please try to listen beyond the distortion and concentrate on identifying the voice. Are the two voices the same or different? You have the option to indicate how sure you are of your decision.''} The scale for judging was: ``Same, absolutely sure'', ``Same, not sure'', ``Different, not sure'' and ``Different, absolutely sure''. The trials consisted of comparisons of VC samples with either the source speaker or the target speaker.
    
The evaluation was designed with crowdsourcing in mind, keeping it modular and scalable in the shape of evaluation sets. Each evaluation set was comprised of 44 utterances to be evaluated: 32 for naturalness and 12 for similarity, all of them corresponding to a total of 4 different systems. As such, each set consisted of 11 samples of a unique system, 8 were rated in terms of naturalness and 3 in terms of similarity (one forced to be cross-gender, another to be same-gender and the other was random). In this fashion, 643 sets provided full coverage of the test samples: 28,292 utterances, corresponding to the 36 evaluated systems (23 Hub systems, 11 Spoke systems, and the sprocket baseline systems, described later, in both tasks) plus the natural speech of the source and target speakers. The target of the evaluation is to rate each sample 4 times, requiring 2,572 sets. This guarantees at least 260 unique evaluators since we limit the maximum of sets per crowdsourced participant to 10.

\section{Evaluation results}

In this section, we will briefly comment on some particularities of the participating teams (subsection~\ref{ssec:participants}), followed by a general description of the baselines provided to the participants (subsection~\ref{ssec:baselines}). Fortunately, team N10 agreed to provide us with a more detailed description of their system to explain how they obtained their impressive results (subsection~\ref{ssec:n10}). The results of the perceptual evaluation are presented in section~\ref{ssec:preceptual} and they are complemented with our analysis of an additional objective measure based on word error rates on the evaluated data (subsection~\ref{ssec:objective}).

\subsection{Challenge Participants}\label{ssec:participants}

Table~\ref{tab:teams} shows a list of the participating teams, their institutional affiliations, and in which tasks they participated. A total of 23 teams submitted systems to the Hub task of the challenge, with 11 of them additionally participating in the Spoke task. Three of the teams came from industry-related companies and the other 20 are composed of university research teams.

Out of the participants, 6 reported having based their work on the Merlin baseline system and 4 on the sprocket baseline system. In terms of vocoders: 11 teams reported having used WORLD, 5 having used sprocket's vocoder-free system, 4 having used STRAIGHT, 2 having used AHOcoder, 2 using Wavenet, 1 using Griffin-Lim, and 1 using IRCAM SuperVP. One team did not submit their basic system information. It must be noted that some teams used different waveform generation methods in the Hub and Spoke tasks and others the same, thus the total number of shown vocoders do not add up to 34.

Additionally, 8 of the teams reported having used long short-term memory recurrent neural network (LSTM-RNN)-based models for their conversion, 4 having used GMMs, 2 having used feed-foward DNNs, 1 having used CycleGAN, and 1 having used deep relational models (DRM) plus adaptive restricted Boltzmann machines (ARBM).

Despite the use of additional data having been allowed in this iteration of the challenge, only 3 teams decided to use that chance to improve their conversion models. Also, the teams that utilized Wavenet as a vocoder reported having used external data for its training.

\begin{table*}[tp]
  \centering
  \caption{Team names and participant institutions of VCC2018. H and S refer to them participating in the Hub and Spoke task respectively.}
    \begin{tabular}{|l|l|l|}\hline
    Team name & Institution name & Tasks \\\hline
    AhoLab & University of the Basque Country & H \\
    AS    & STMS-IRCAM/Sorbonne University/CNRS/IntelligentVoice & H,S \\
    AST   & Academia Sinica & H,S \\
    Azurite & Indian Institute of Technology Bombay & H,S \\
    CMU   & Carnegie Mellon University & H \\
    CPqD  & CPqD  & H \\
    CSLU  & Oregon Health \& Science University & H \\
    CSTR  & University of Edinburgh & H \\
    CUHK  & The Chinese University of Hong Kong & H,S \\
    DA-IICT & Dhirubhai Ambani Institute of Information and Communication Technology & H,S \\
    DSP-AGH & AGH University of Sciencie and Technology & H \\
    Hulk2 & Shanghai Jiao Tong University & H \\
    NWPU-I2R-NUS & Northwestern Polytechnical University/Institute for Infocomm Research/National University of Singapore & H \\
    NTT-CSlab & Nippon Telegraph and Telephone Corporation & H,S \\
    NTU   & Nanyang Technological University & H,S \\
    NTUT  & National Taipei University of Technology & H \\
    NU    & Nagoya University & H,S \\
    PDL   & Pindrop & H \\
    RBM   & University of Electro-Communications & H,S \\
    TEXAGS & Texas A\&M University & H,S \\
    USTC  & University of Science and Technology of China & H,S \\
    UTokyo & The University of Tokyo & H \\
    xmuspeech & Xiamen University & H \\\hline
    \end{tabular}%
  \label{tab:teams}%
  \vspace{-3mm}
\end{table*}%

\subsection{Baseline system}\label{ssec:baselines}

For this challenge, we released two baseline systems that could be used by the participants to base their work on and to provide an easier starting ground to newcomers to the tasks: sprocket (B01 in the evaluation) and Merlin (not evaluated). sprocket is an open source implementation of the GMM- and differential GMM (DIFFGMM)-based systems \cite{sproket}\footnote{\url{https://github.com/k2kobayashi/sprocket}}. Merlin is another open source toolkit for building deep neural network models for statistical parametric speech synthesis and voice conversion\cite{wu2016merlin}\footnote{\url{https://github.com/CSTR-Edinburgh/merlin}}. We decided to provide both environments to cover all of the following: a DNN-based approach, a traditional VC system based on a GMM \cite{toda2007voice}, and a vocoder-free VC system based on a DIFFGMM \cite{Kobayashi2016}.  

For the baseline system B01, vocoder-free and vocoder-based VC systems were respectively developed for the same-gender and cross-gender conversion pairs using sprocket. For the Spoke task, gender-dependent many-to-one VC systems were developed using only the dataset of the Hub task. More specifically, using two source male speakers and a single target male speaker in the Hub task, a gender-dependent, target-specific many-to-one VC system was developed and used for converting the different source male speakers in the Spoke task into the target male speaker. We built similar many-to-one VC systems for the rest of the target speakers and likewise used them in the Spoke task. 

\subsection{The N10 System}\label{ssec:n10}

In N10, a speaker-independent content-posterior-feature extractor was first built using hundreds of hours of external speech data with aligned phonetic transcriptions. This model extracted content posterior features from source speech at the conversion stage. Then, a speaker-dependent LSTM-RNN was used to predict F0 and STRAIGHT spectral features \cite{kawahara99straight} from the content posterior features for each target speaker. Finally, a speaker-dependent WaveNet vocoder \cite{tamamori2017speaker, hu2017ustc} was built to reconstruct the waveforms from the predicted F0 and spectral features. Some manual checks and annotations of the training data were also conducted. These include corrections of F0 extraction errors and removals of some speech segments with irregular phonation. The same framework was adopted for both Hub and Spoke tasks.

\subsection{Perceptual evaluation}\label{ssec:preceptual}

We carried out a crowdsourced perceptual evaluation as explained in section~\ref{subsec:eval}. A total of 267 unique listeners (146 male and 121 female) completed the evaluation, with an average of 9.4 sets completed per participant, for a total of 113,168 evaluated utterances in 2,572 sets. 

\subsubsection{Hub task results}

\begin{figure}[tb]
        \centering
        \includegraphics[width=8cm]{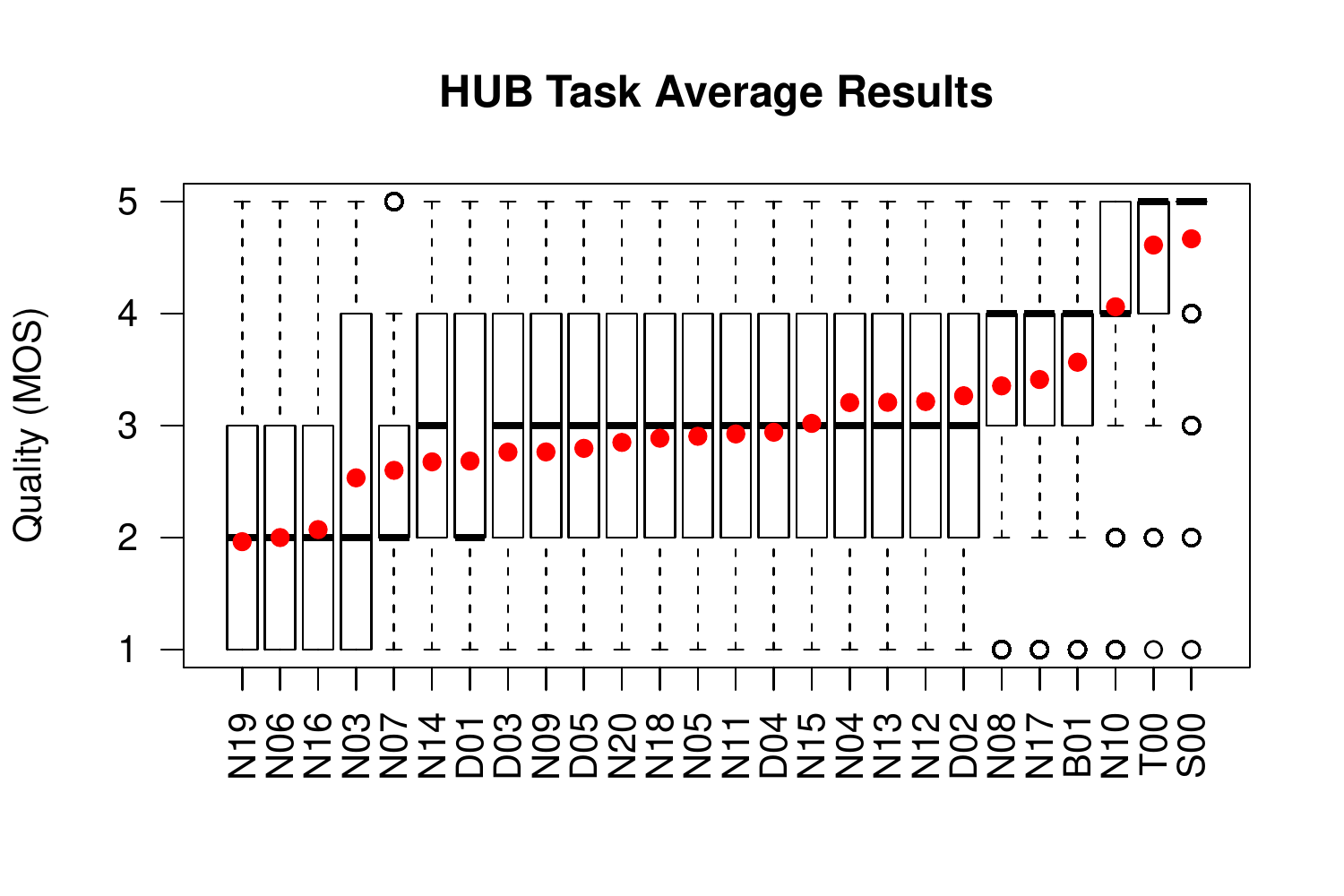}
        \vspace{-5mm}
        \caption{Naturalness results of the Hub task for all speaker pairs. MOS scores are averaged across all pairs, arranged in accordance with their mean (red dot).}
        \label{fig:nat_hub}
        \vspace{-5mm}
      \end{figure}

\begin{figure}[tb]
        \centering
        \includegraphics[width=8cm]{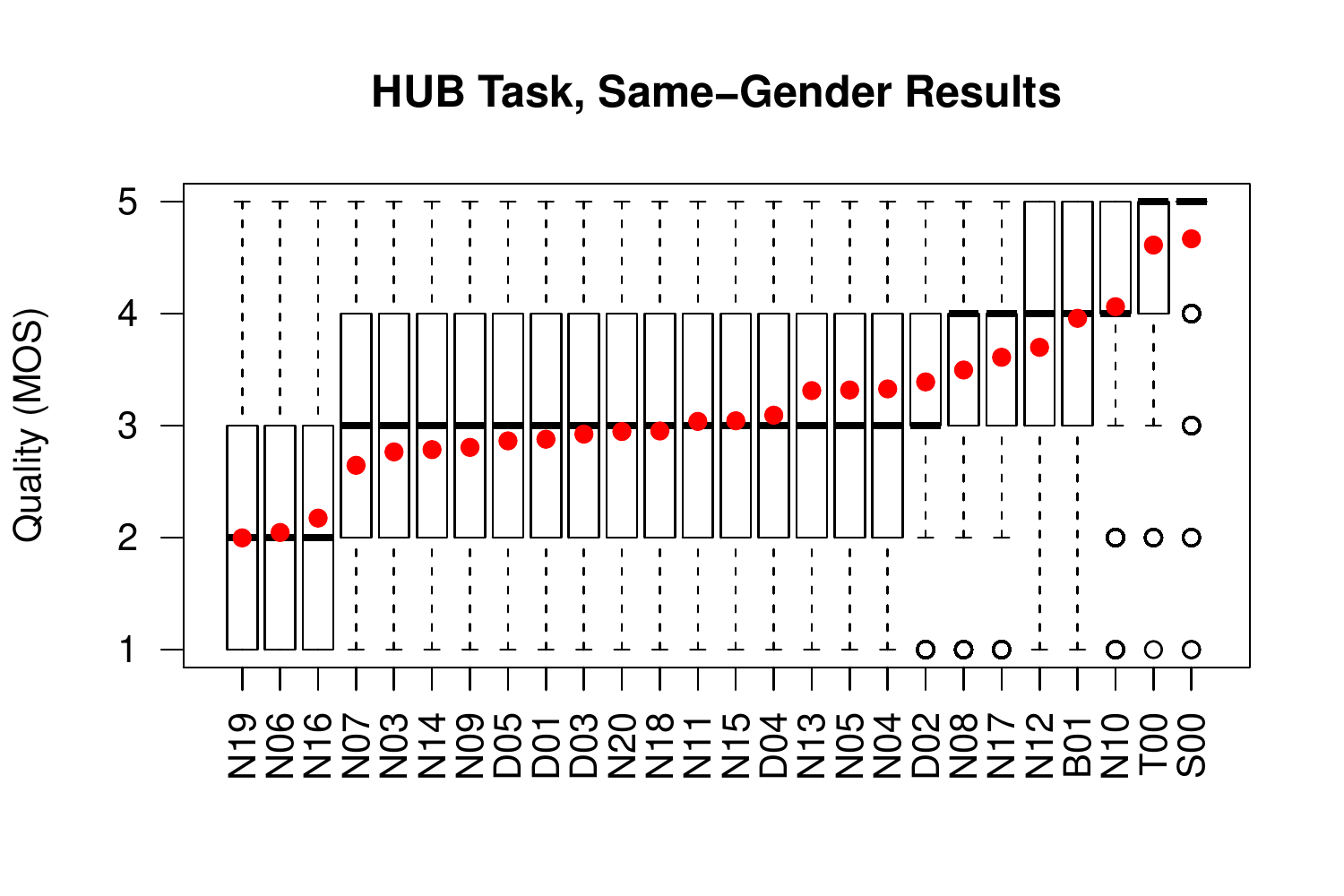}
        \vspace{-5mm}
        \caption{Naturalness results of the Hub task for same-gender conversion pairs. MOS scores are averaged across all pairs, arranged in accordance with their mean (red dot).}
        \label{fig:nat_hub_sg}
        \vspace{-5mm}
      \end{figure}

\begin{figure}[tb]
        \centering
        \includegraphics[width=8cm]{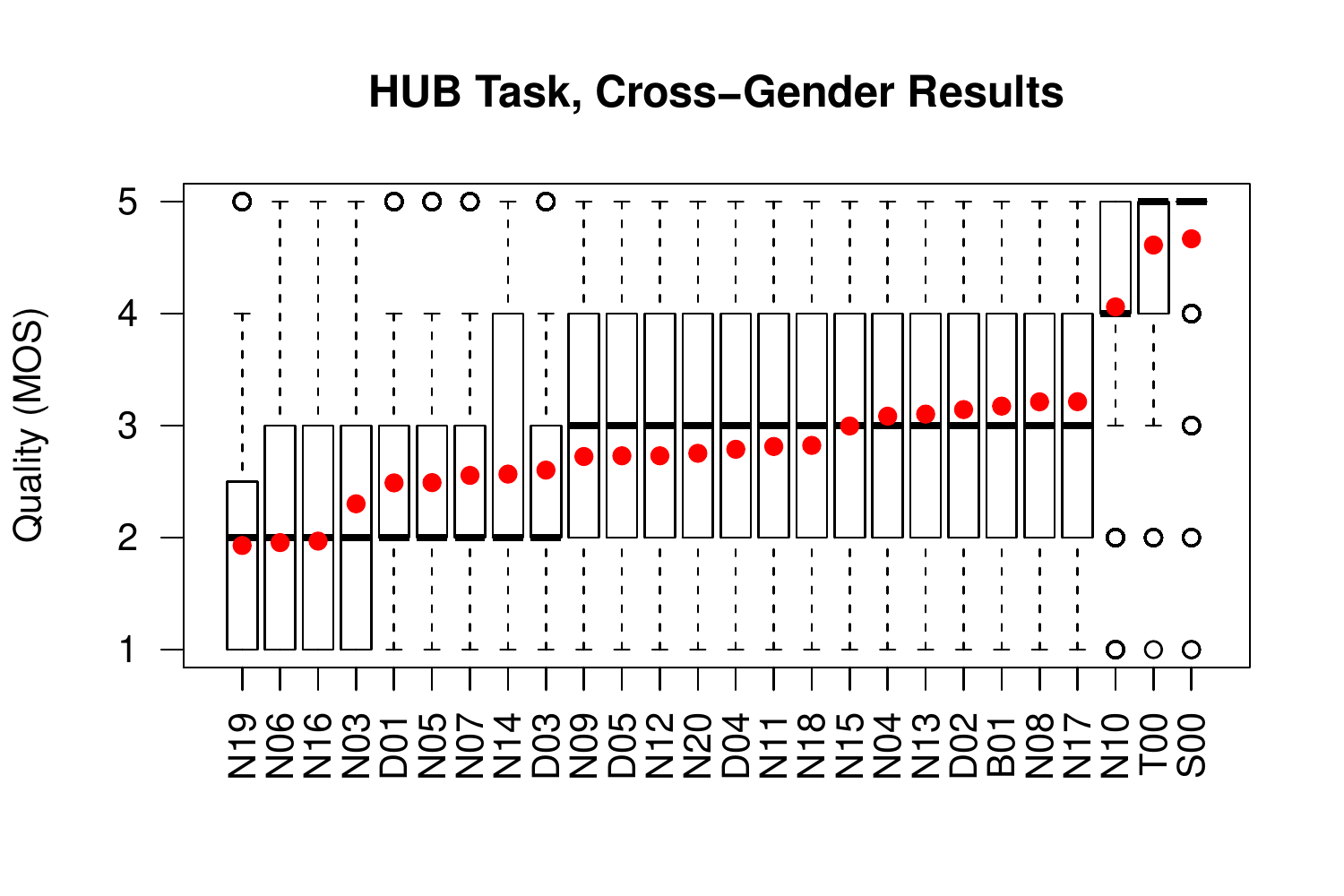}
        \vspace{-5mm}
        \caption{Naturalness results of the Hub task for cross-gender speaker pairs. MOS scores are averaged across all pairs, arranged in accordance with their mean (red dot).}
        \label{fig:nat_hub_xg}
      \end{figure}

Figures~\ref{fig:nat_hub}, \ref{fig:nat_hub_sg}, and \ref{fig:nat_hub_xg} show the boxplots for the naturalness evaluation results of the Hub task when considering all speaker pairs, the same-gender conversion pairs, and the cross-gender pairs, respectively. S00 refers to the natural speech of the source speaker, T00 to the natural speech of the target speaker, and B01 to the sprocket baseline.

The sprocket-based baseline outperforms the majority of the systems in terms of naturalness, with only N10 and the natural speech of the source and target speakers outperforming it significantly. 

      
A similar trend can be observed from the same-gender conversion pair results, where N10 outperforms all the other presented systems, and the baseline shows very competitive results.

However in the cross-gender results, although N10 is clearly the top-performing system without any significant reduction in speech naturalness, all other systems show a significant drop in average naturalness when compared to their same-gender results (2.78 vs. 3.08, a 0.3 drop in mean opinion score).

\begin{figure}[tb]
        \centering
        \includegraphics[width=8cm]{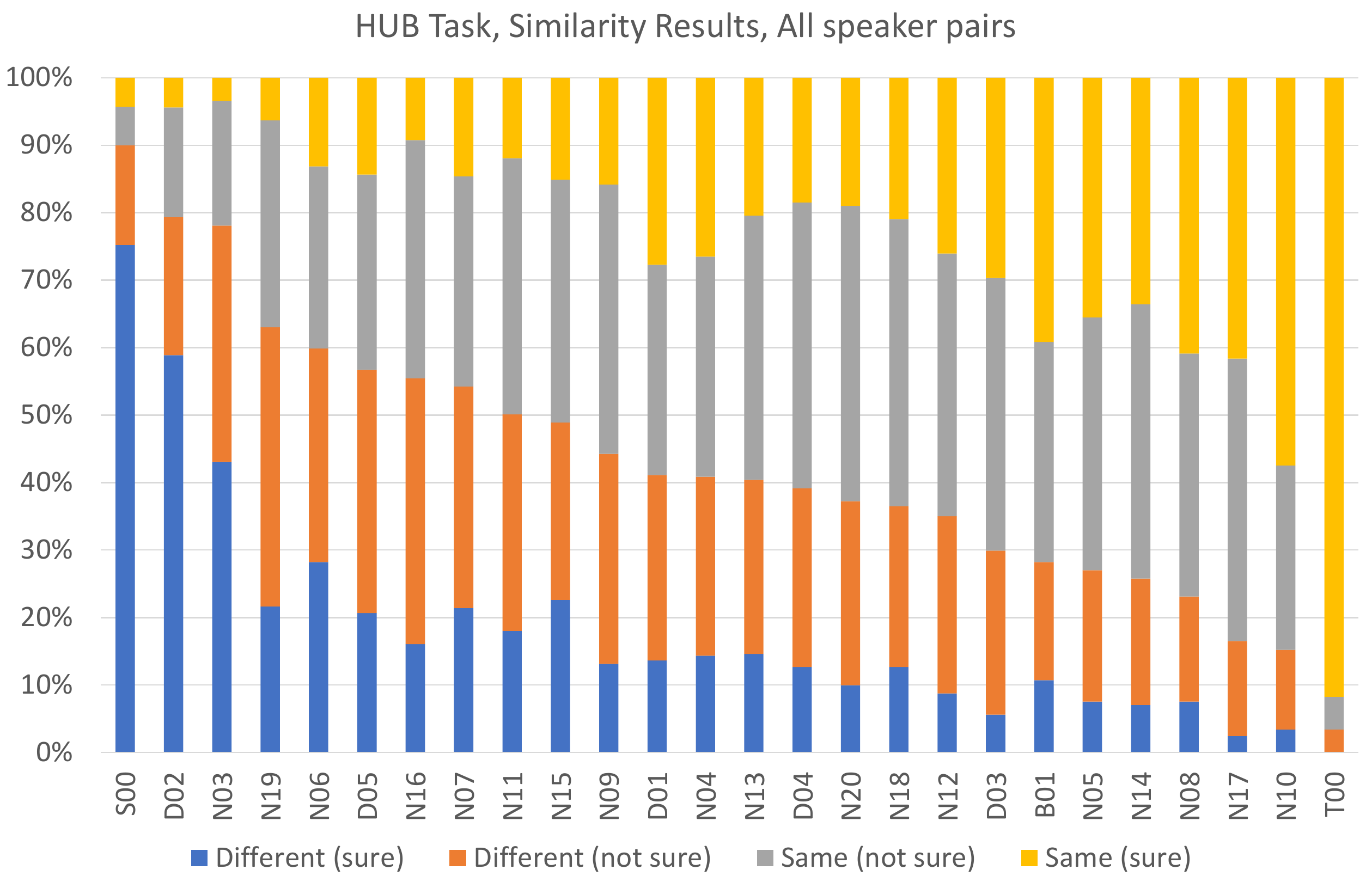}
        \caption{Similarity results of the target speaker for the Hub task averaged across all speaker pairs.}
        \label{fig:sim_hub}
        \vspace{-5mm}
      \end{figure}

The similarity evaluation results (figure~\ref{fig:sim_hub}) show that N10 is still the best performing system with a similarity score very close to that of the target natural speech, but is contested by N17, which provides comparable similarity results. The results also show that even though a number of teams outperformed the baseline, they still provided a competitive result. In general, most teams managed to successfully convert the identity of the speaker with similarity scores higher than 50\%. The similarity score is defined as the added percentage of same (not sure) and same (sure) scores for the system. 

\begin{figure}[tb]
        \centering
        \includegraphics[width=8cm]{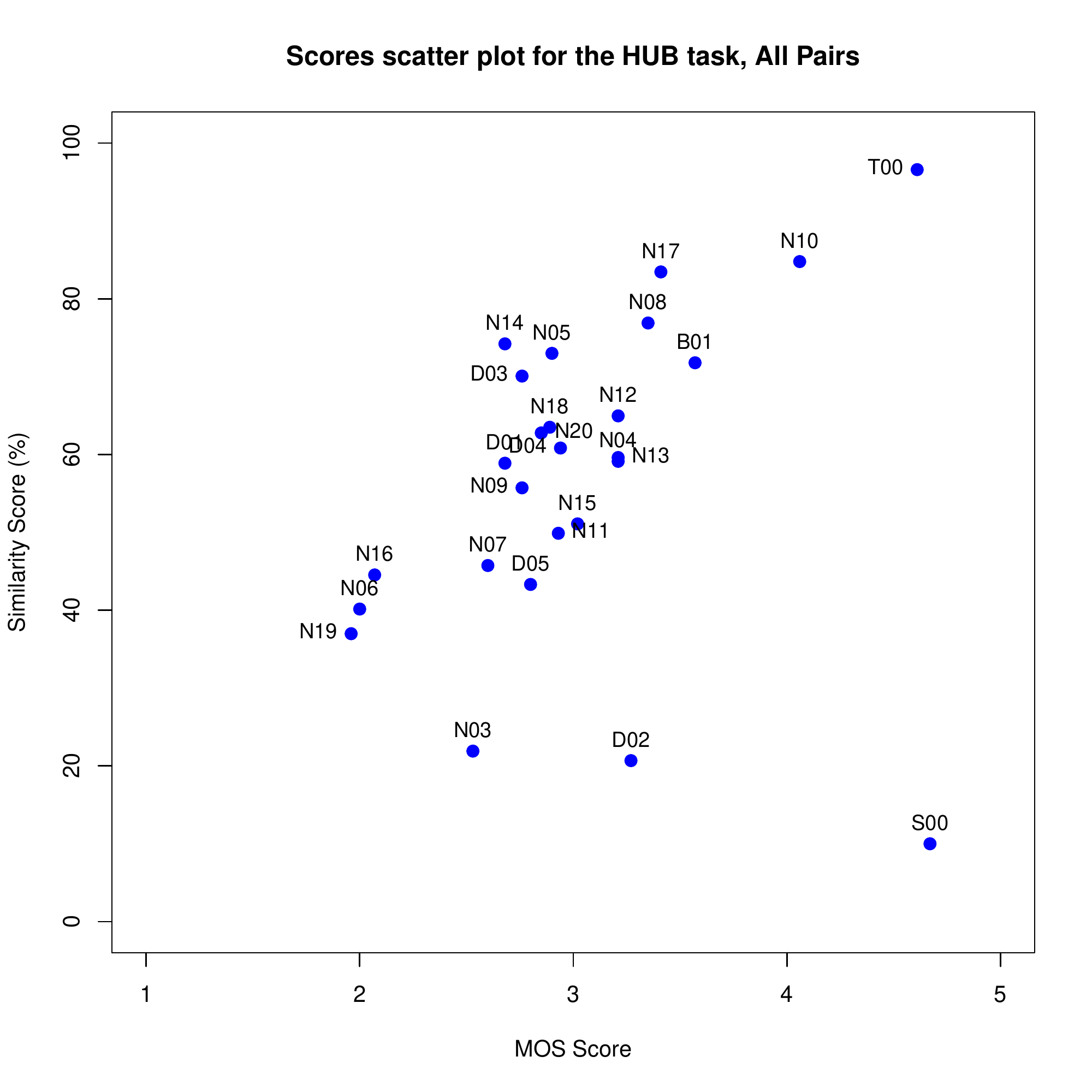}
        \caption{Scatter plot matching naturalness and similarity scores to target speaker for the Hub task when averaging all speaker pairs.}
        \label{fig:scatter_hub}
        \vspace{-5mm}
      \end{figure}

Figure~\ref{fig:scatter_hub} shows a scatter-plot matching naturalness and similarity scores to the target speaker for the Hub task when averaging all speaker pairs. This figure enables us to easily compare the trade-offs that most systems have to do to improve either similarity or naturalness. N10 is again the closest to the actual natural speech of the target T00, but N17 also separates itself from the pack thanks to its competitive similarity and naturalness results. It should be noted that both N10 and N17 used the Wavenet vocoder.

\subsubsection{Spoke task results}

\begin{figure}[tb]
        \centering
        \includegraphics[width=8cm]{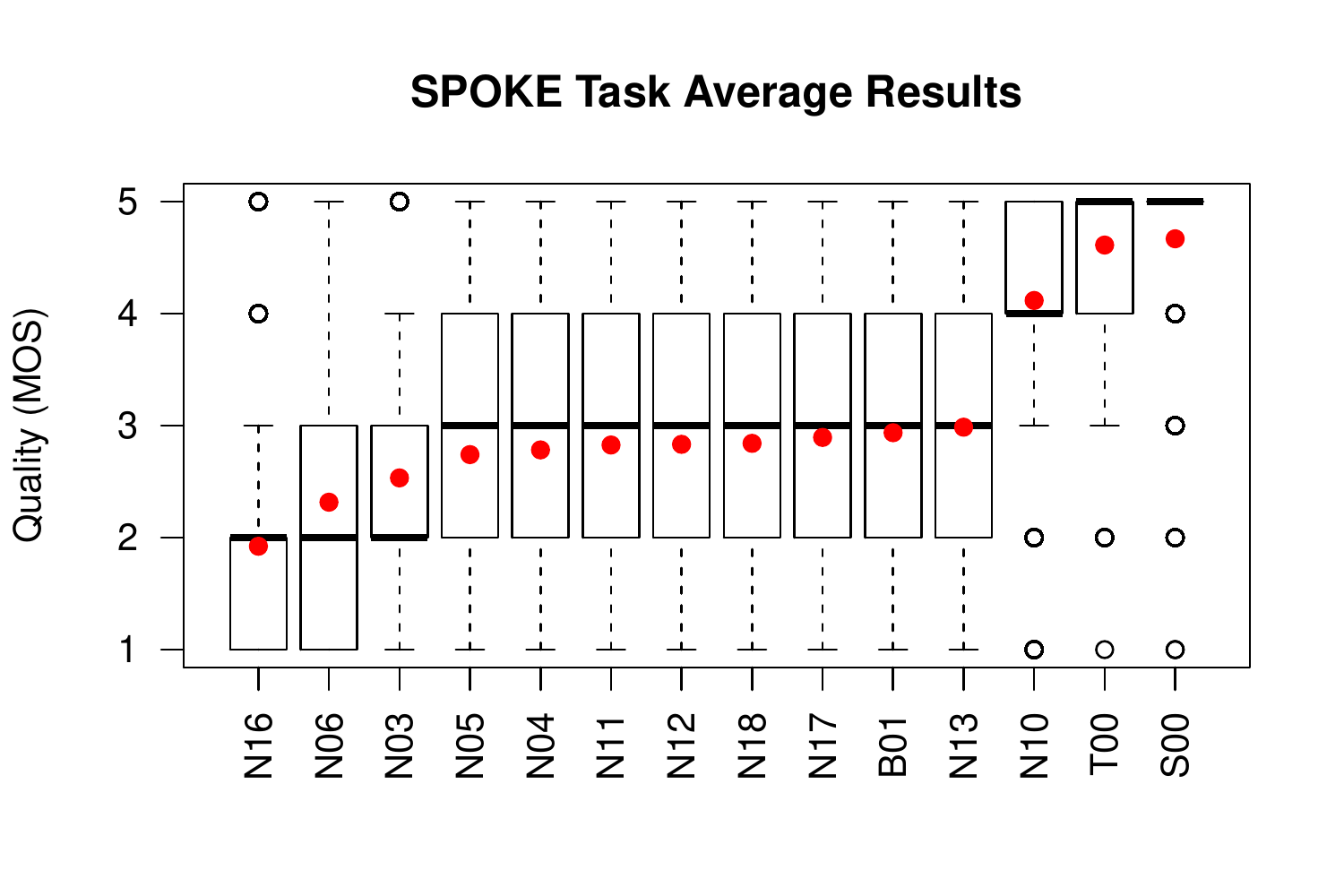}
        \vspace{-5mm}
        \caption{Naturalness results of the Spoke task for all speaker pairs. MOS scores are averaged across all pairs, arranged in accordance with their mean (red dot).}
        \label{fig:nat_spo}
        \vspace{-5mm}
      \end{figure}

\begin{figure}[tb]
        \centering
        \includegraphics[width=8cm]{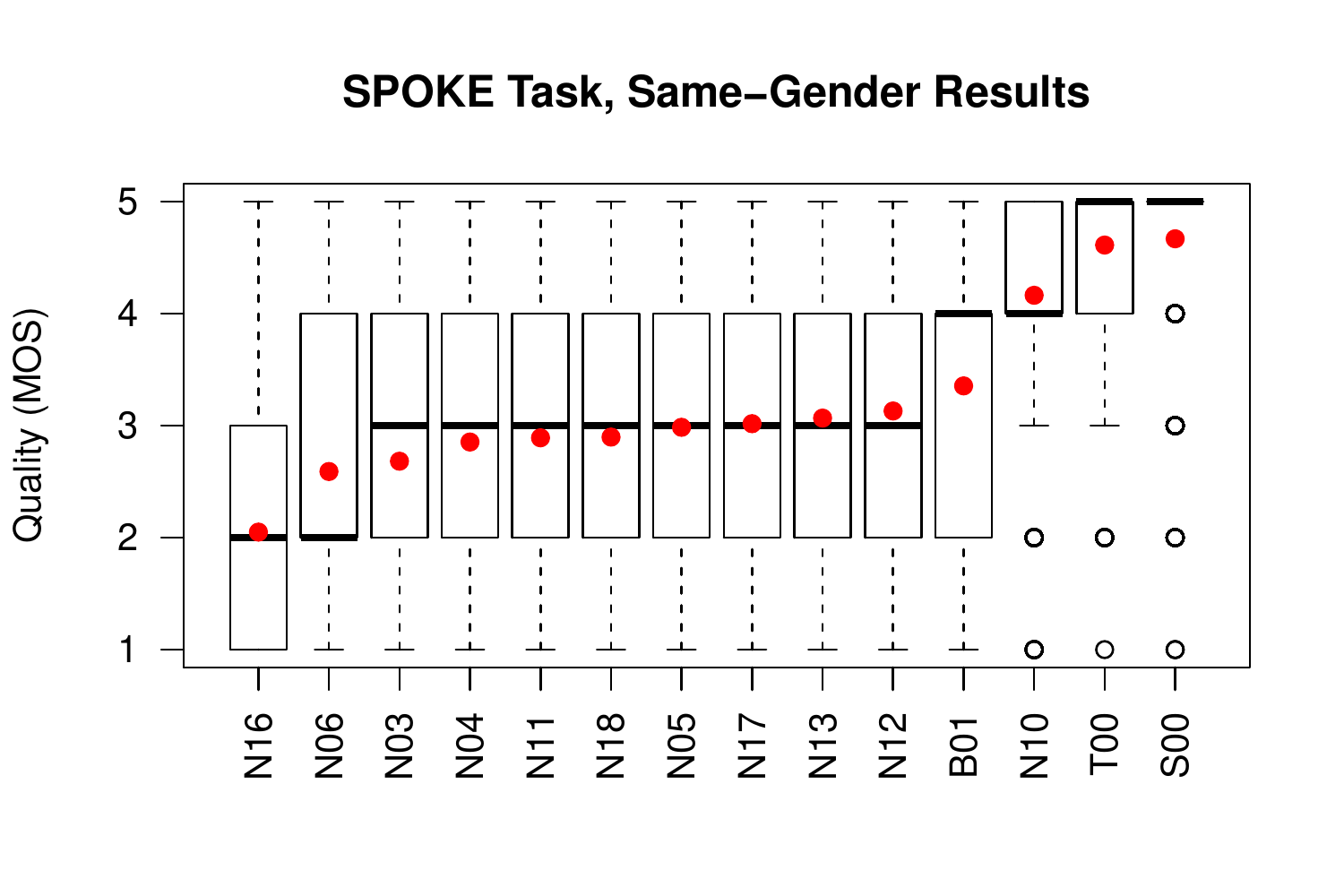}
        \vspace{-5mm}
        \caption{Naturalness results of the Spoke task for same-gender conversion pairs. MOS scores are averaged across all pairs, arranged in accordance with their mean (red dot).}
        \label{fig:nat_spo_sg}
        \vspace{-5mm}        
      \end{figure}

\begin{figure}[tb]
        \centering
        \includegraphics[width=8cm]{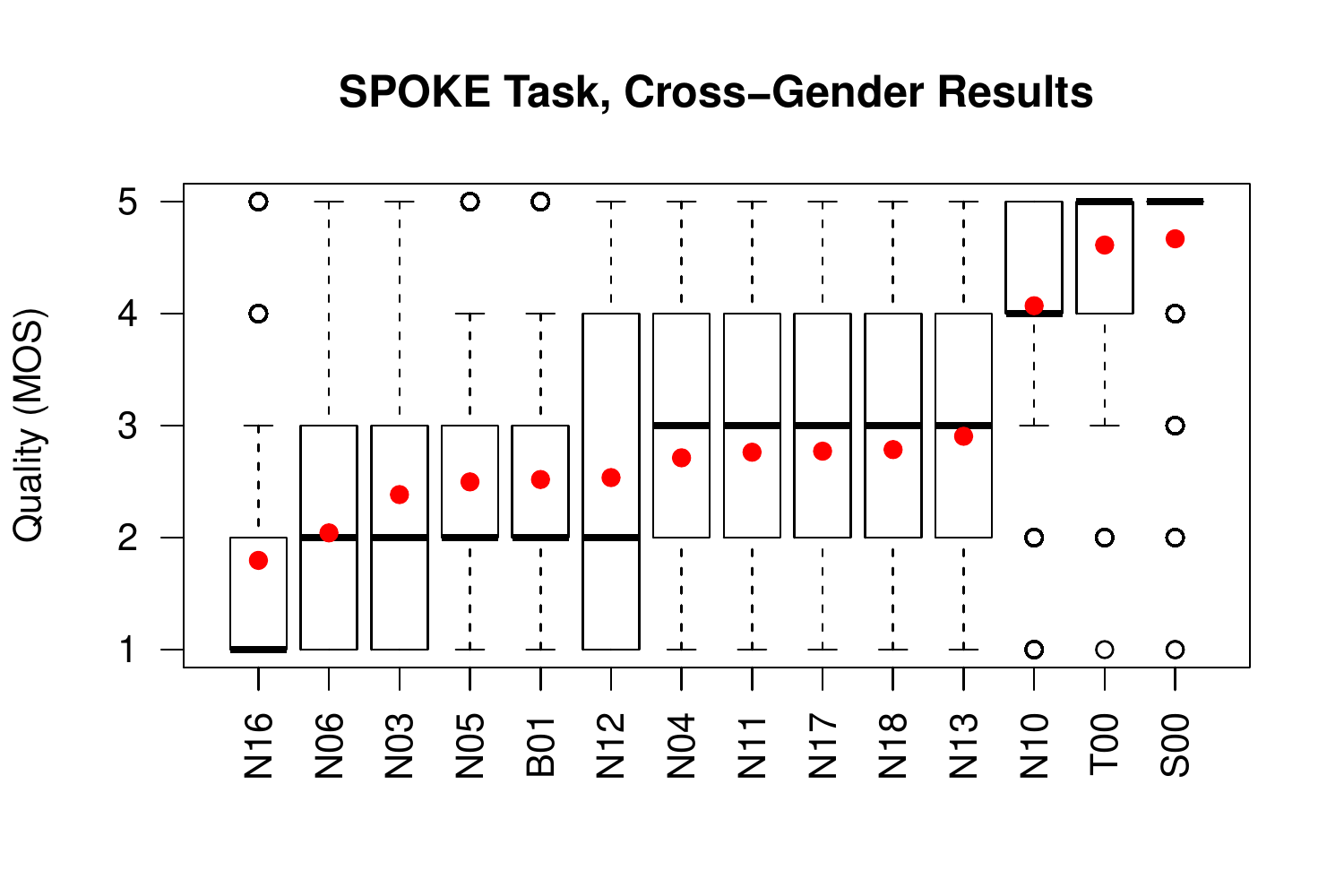}
        \vspace{-5mm}
        \caption{Naturalness results of the Spoke task for cross-gender speaker pairs. MOS scores are averaged across all pairs, arranged in accordance with their mean (red dot).}
        \label{fig:nat_spo_xg}
        \vspace{-5mm}
      \end{figure}

Figures~\ref{fig:nat_spo}, \ref{fig:nat_spo_sg}, and \ref{fig:nat_spo_xg} show the boxplots for the naturalness evaluation results of the Spoke task when considering all speaker pairs, the same-gender conversion pairs, and the cross-gender pairs, respectively. Compared to that of the Hub task results, there is an average drop in naturalness of 0.09 points in the MOS score, 0.17 if we exclude N10 from the calculation. This result clearly reflects the increase in complexity between the parallel and non-parallel conversion task.

By analyzing the same-gender and cross-gender pairs component of the results, we see similar pattern to that in the Hub task. N10 is very stable regardless of the conversion conditions. The baseline performs very well in the same-gender case, outperforming all the other systems, but falls significantly when considering the cross-gender case. On average, we observed a drop of 0.36 points in the MOS score from the same-gender (2.95) to the cross-gender condition (2.59) if we exclude N10. 

\begin{figure}[tb]
        \centering
        \includegraphics[width=8cm]{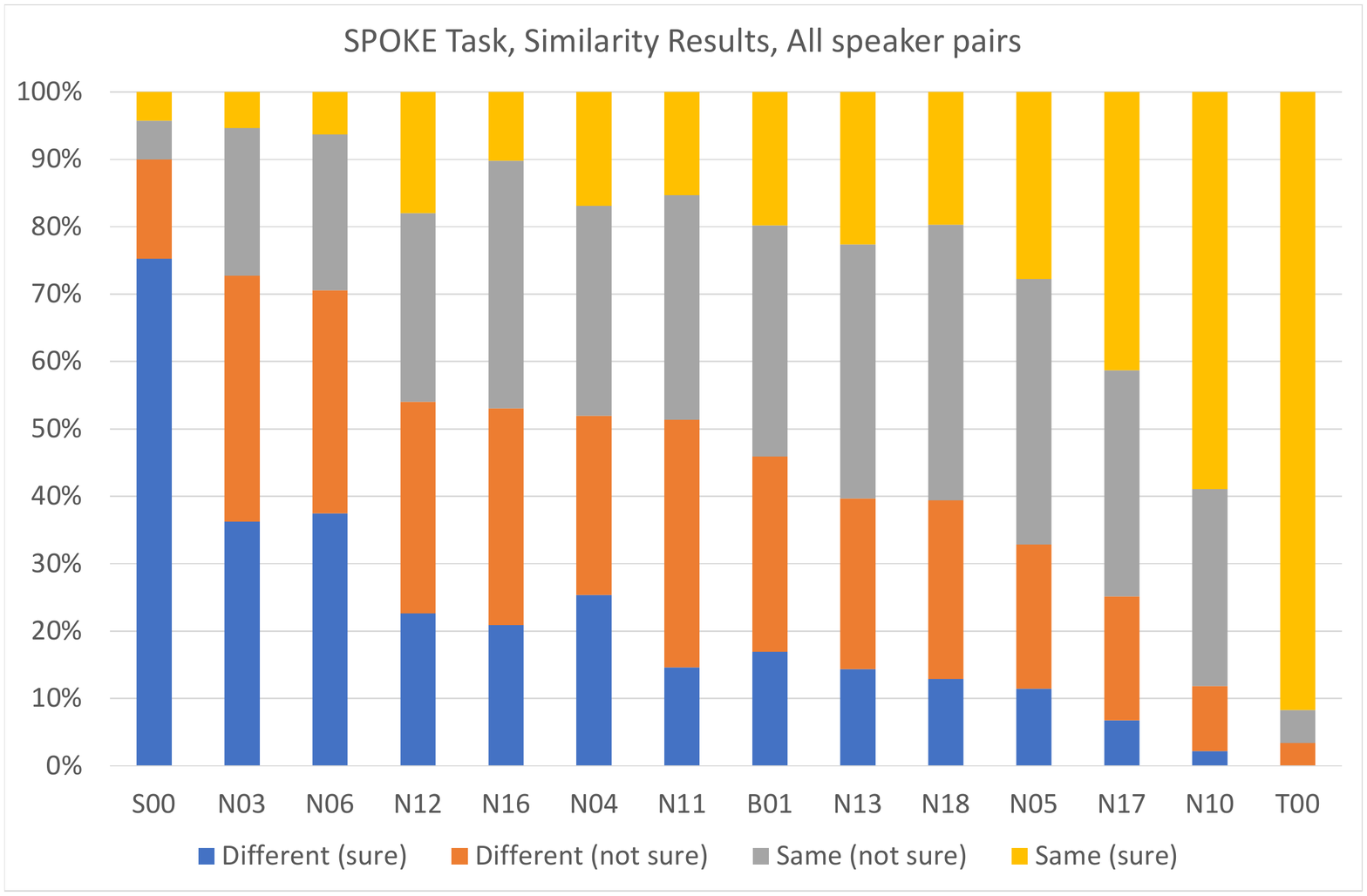}
        \vspace{-4mm}
        \caption{Similarity results of the target speaker for the Spoke task averaged across all speaker pairs.}
        \label{fig:sim_spo}
      \end{figure}

The similarity evaluation results for the Spoke task (Figure~\ref{fig:sim_spo}) also show some encouraging results for the non-parallel conversion task. Not only did N10 manage to successfully imbue the target speaker identity despite the challenging task, but 5 other systems also managed to convert the speech with similarity scores higher than 50\% (including the baseline).

\begin{figure}[tb]
        \centering
        \includegraphics[width=8cm]{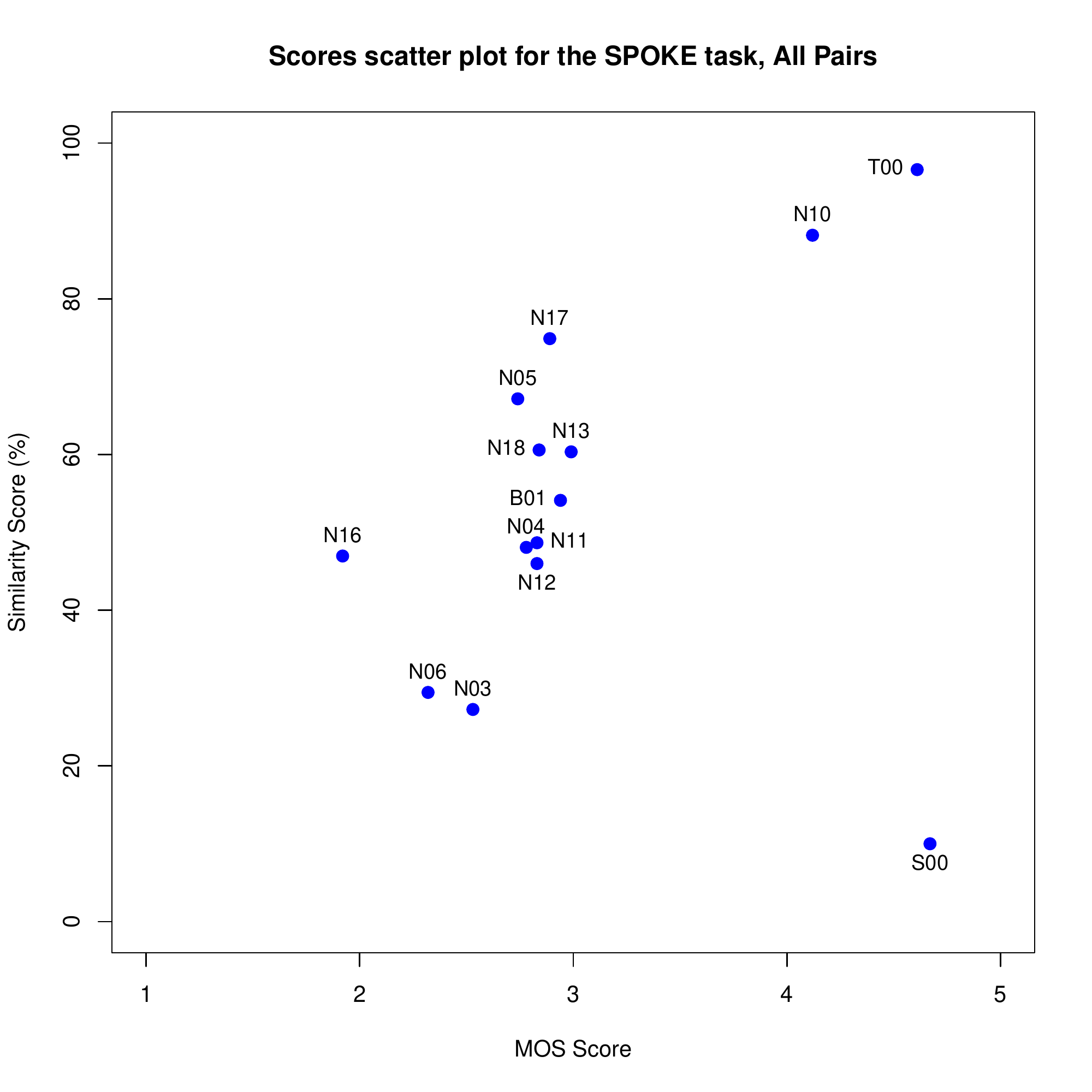}
        \vspace{-5mm}
        \caption{Scatter plot matching naturalness scores and similarity to target speaker for the Spoke task  when averaging all speaker pairs.}
        \label{fig:scatter_spo}
        \vspace{-5mm}
      \end{figure}

Figure~\ref{fig:scatter_spo} shows a scatter plot matching naturalness scores and similarity to the target speaker for the Spoke task in this case. In this evaluation, the supremacy of N10 over the rest of the submitted systems becomes more evident, keeping its performance very close to that of the actual target speaker. The remaining systems average a MOS score of 3 with different successes in terms of similarity, none coming close to their Hub task performance.

\subsection{Word error rate}\label{ssec:objective}

In VCC 2018, no subjective listening tests on the intelligibility of the converted speech were conducted. To roughly evaluate the linguistic consistency after voice conversion, the word error rates (WERs) of transcribing the converted speech using an automatic speech recognition (ASR) engine were calculated for all entries and all source-target pairs. The ASR engine was a prototype system developed by iFlytek, which adopted a state-of-the-art neural-network-based ASR architecture and was trained using 10,000hrs-level recordings for acoustic modeling and GB-level texts for language modeling. The vocabulary size was around 200,000. WERs were calculated by the \emph{HResults} tool in HTK \cite{young2002htk} using manual transcriptions as ground truth. The WERs of all systems on the Hub and Spoke tasks are summarized in Table \ref{tab:wer}.

\begin{table}[tb]
  \centering
  \caption{WERs (\%) of all systems in the Hub and Spoke tasks, where ``B01" is the baseline system and ``source" denotes the natural speech from source speakers.}
    \label{tab:wer}

    \begin{tabular}{c c c | c c c}
        \hline
        \textbf{System}  & \textbf{Hub} & \textbf{Spoke} & \textbf{System}  & \textbf{Hub} & \textbf{Spoke} \\
        \hline
        B01 &  10.47 &  20.11 & N09 &  12.45 &   -   \\
        D01 &  18.51 &   -    & N10 &  10.30 &  9.13  \\
        D02 &  8.74  &   -    & N11 &  16.34 &  16.68 \\
        D03 &  19.27 &   -    & N12 &  18.43 &  26.17 \\
        D04 &  23.54 &   -    & N13 &  16.46 &  19.58 \\
        D05 &  15.75 &   -    & N14 &  17.85 &   -    \\
        N03 &  20.12 &  22.41 & N15 &  13.72 &   -   \\
        N04 &  14.02 &  13.60 & N16 &  35.60 &  34.99\\
        N05 &  26.51 &  26.24 & N17 &  17.80 &  26.86\\
        N06 &  36.47 &  23.85 & N18 &  8.60  &  9.04 \\
        N07 &  17.11 &   -    & N19 &  19.73 &   -    \\
        N08 &  11.74 &   -    & N20 &  13.75 &   -    \\
        source &  6.40  &   -   & & &\\
        \hline
    \end{tabular}
\end{table}

The natural speech of the source speakers achieved a lower WER than that for all converted speech. It is reasonable since the signal processing procedures during voice conversion introduced spectral distortions into natural speech and increased the difficulty of ASR. Comparing the WERs of one system in both tasks, there was no clear indication that using non-parallel data may lead to higher WERs. For the Hub task, the correlation coefficients between the WERs of all systems and the MOS and similarity scores were $-0.6587$ and $-0.3218$, respectively. For the Spoke task, they were $-0.7127$ and $-0.2272$, respectively. These results indicate that there was a strong negative correlation between MOS scores and WERs. The reason may be that the spectral distortions of voice conversion degraded the subjective quality of converted speech and decreased the accuracy of ASR simultaneously.

\section{Conclusion}

This paper presented the second edition of the Voice Conversion Challenge (VCC 2018), which has continued the trend of providing a common framework for the development and evaluation of voice conversion systems. In this challenge, we have seen the incredible progress that has come to the field with the rise of new speech generation paradigms such as Wavenet, showing performances that produce converted speech with a quality similar to that of natural speech. From the listening test, we observed that one of the submitted VC systems achieved remarkable results. This system obtained an average of 4.1 in the five-point scale evaluation for quality judgment and about 80\% of its converted speech samples were judged to be the same as target speakers by listeners. We see the results of the VCC 2018 as a potential paradigm shift in the field to convince teams all over the world to consider these new approaches. All the training and evaluation data released to participants, submissions from participants, and the listening test results are publicly and permanently available at the Edinburgh datashare\footnote{http://dx.doi.org/10.7488/ds/2337}.

Analysis of the spoofing performance of these VC systems is described in \cite{tomivcc2analysis}. We determined that these listening test results do not directly reflect spoofing capability.

\noindent\textbf{Acknowledgements:}
We are grateful to iFlytek Ltd.\ for sponsoring the evaluation of the VCC 2018. This work was partially supported by MEXT KAKENHI Grant Numbers (15H01686, 16H06302, 17H04687, 17H06101) and by Academy of Finland (Proj.\ No. 309629).

\bibliographystyle{IEEEtran}
\bibliography{mybib}

\newpage 

\appendix

\section{Breakdown of listeners}

Table~\ref{tab:subjects} gives a breakdown of the age categories and (self-reported) accents of our subjects.

\begin{table}[th]
        \centering
        \caption{{\it Age and accent of subjects.}}
        \label{tab:subjects}
        \begin{tabular}{c|r|l|r}
        \hline
             Age & \# & Accent &  \#\\
             \hline
             18-30 &  116 & North American & 141 \\
             31-40 &  94 & British & 58\\
             41-50 &  45 & Other & 22\\
             51+   &  12 & Non-native: & 46\\
        \hline
        \end{tabular}
\end{table}

\section{Similarity to source speakers}
\begin{figure}[htb]
        \centering
        \includegraphics[width=9cm]{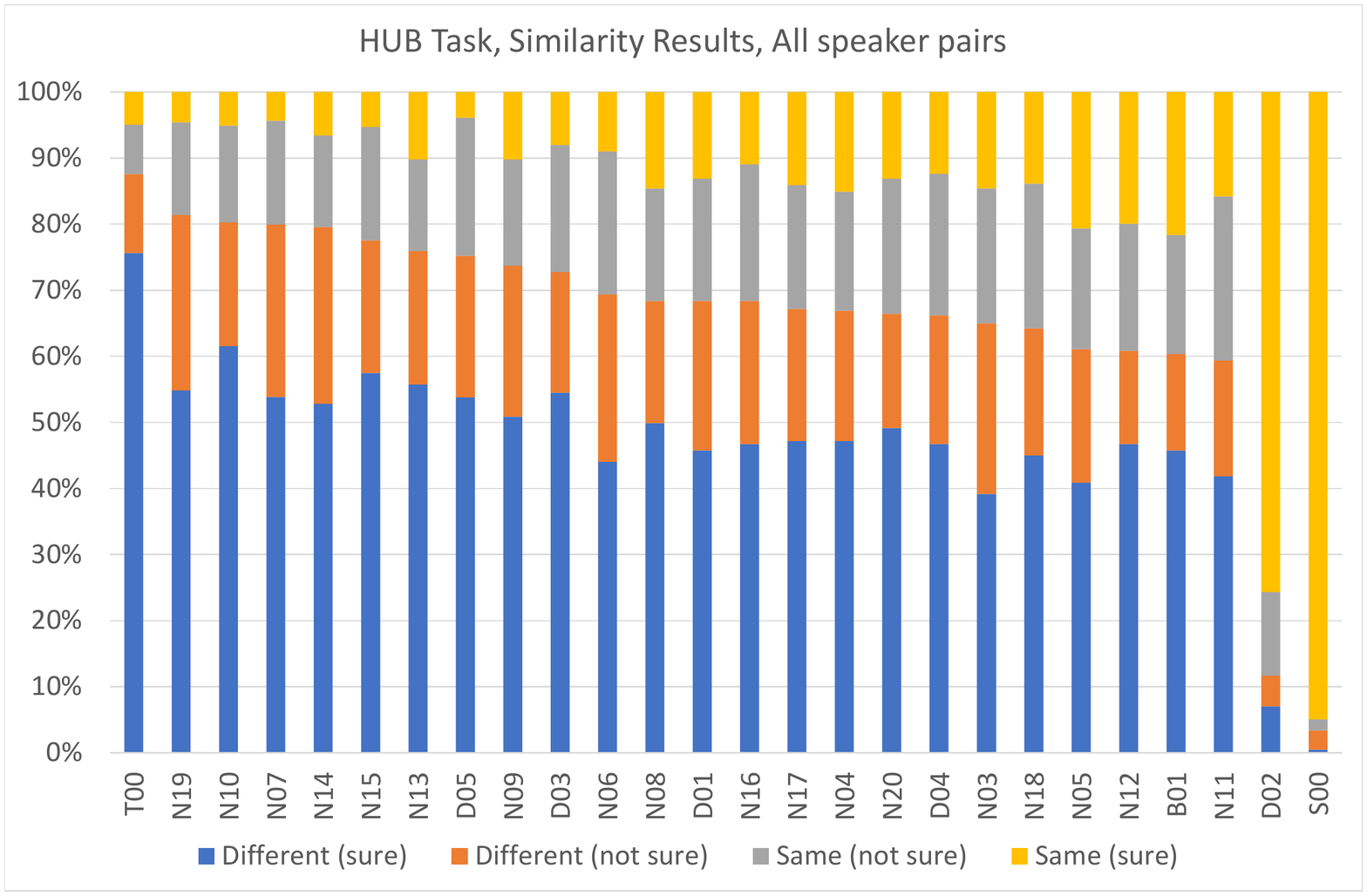}
        \caption{Similarity results of the target speaker for the Hub task averaged across all speaker pairs.}
        \label{fig:sim_hub_src}
\end{figure}

\begin{figure}[htb]
        \centering
        \includegraphics[width=9cm]{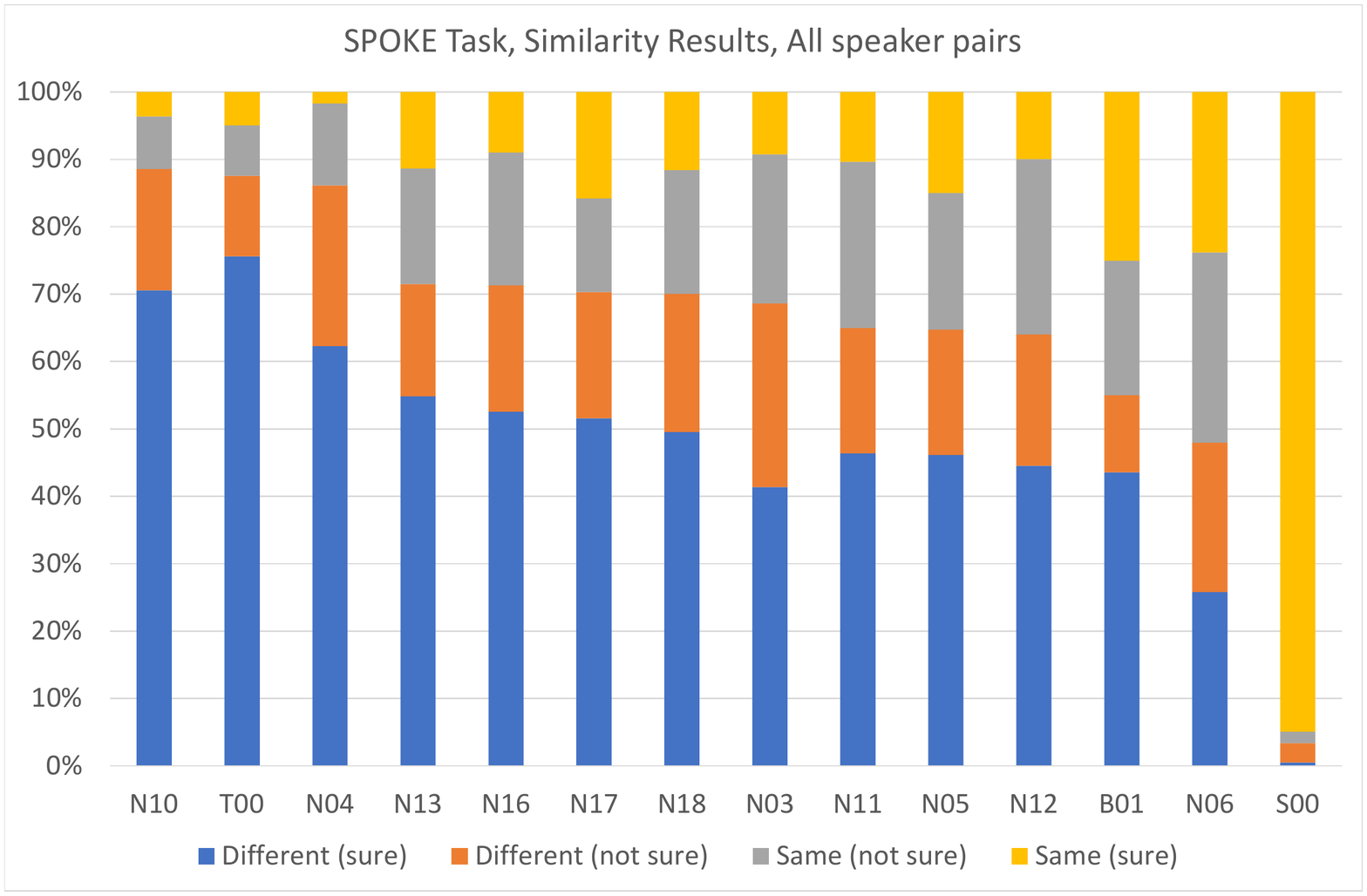}
        \caption{Similarity results of the target speaker for the Hub task averaged across all speaker pairs.}
        \label{fig:sim_spo_src}
\end{figure}

In the listening test, we also measured the similarity of the VC samples to source speakers using the Same/Different paradigm as described earlier. Figures \ref{fig:sim_hub_src} and \ref{fig:sim_spo_src} show the similarity evaluation results to the source speakers. Almost all VC samples apart from D02 were judged as different speakers from the source speakers. 

\section{Scatter plots of ASR WER and MOS}  

\begin{figure}[htb]
        \centering
        \includegraphics[width=8cm]{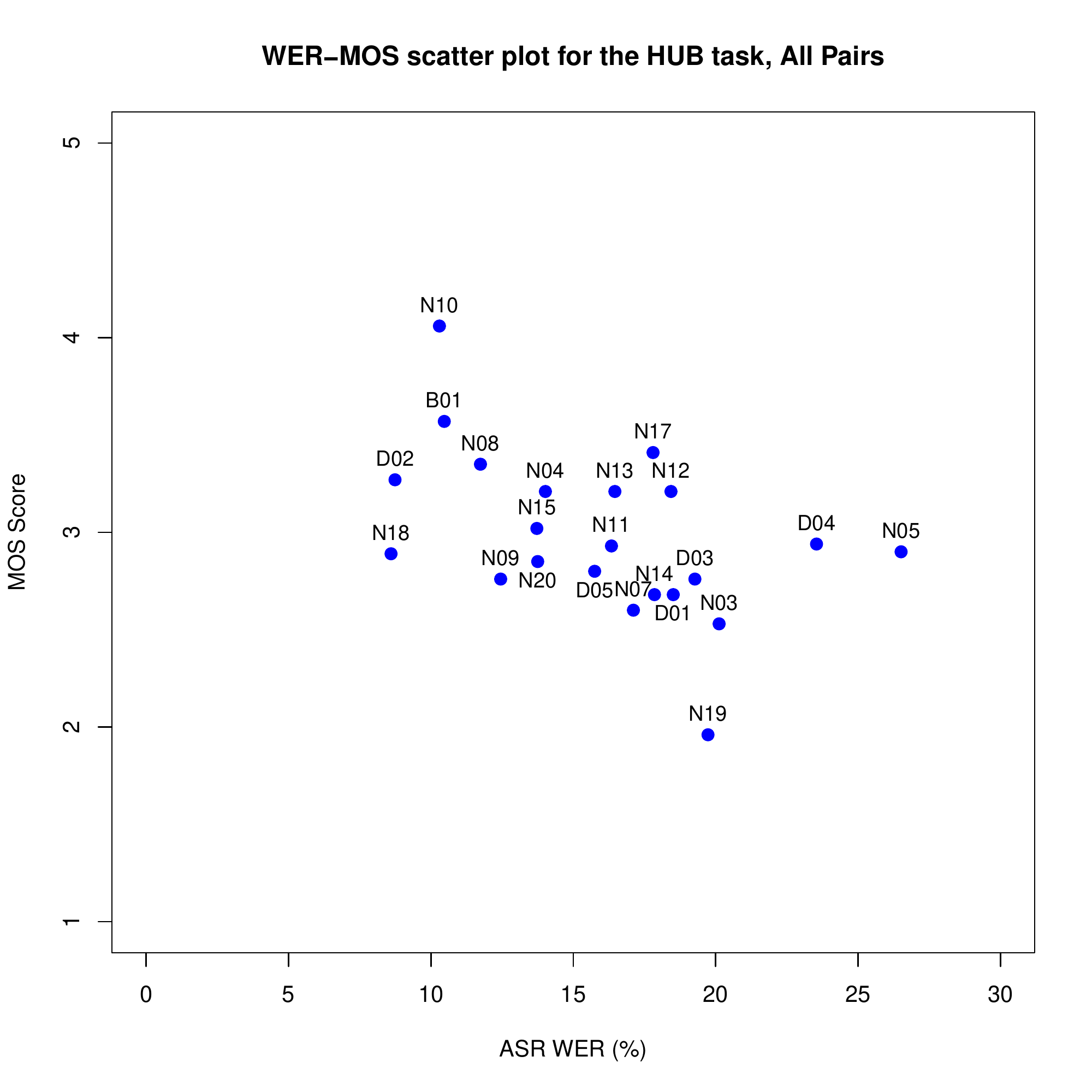}
        \vspace{-5mm}
        \caption{Scatter plot matching naturalness scores and WERs for the Hub task  when averaging all speaker pairs.}
        \label{fig:scatter_wer_hub}
        \vspace{-5mm}
      \end{figure}

\begin{figure}[htb]
        \centering
        \includegraphics[width=8cm]{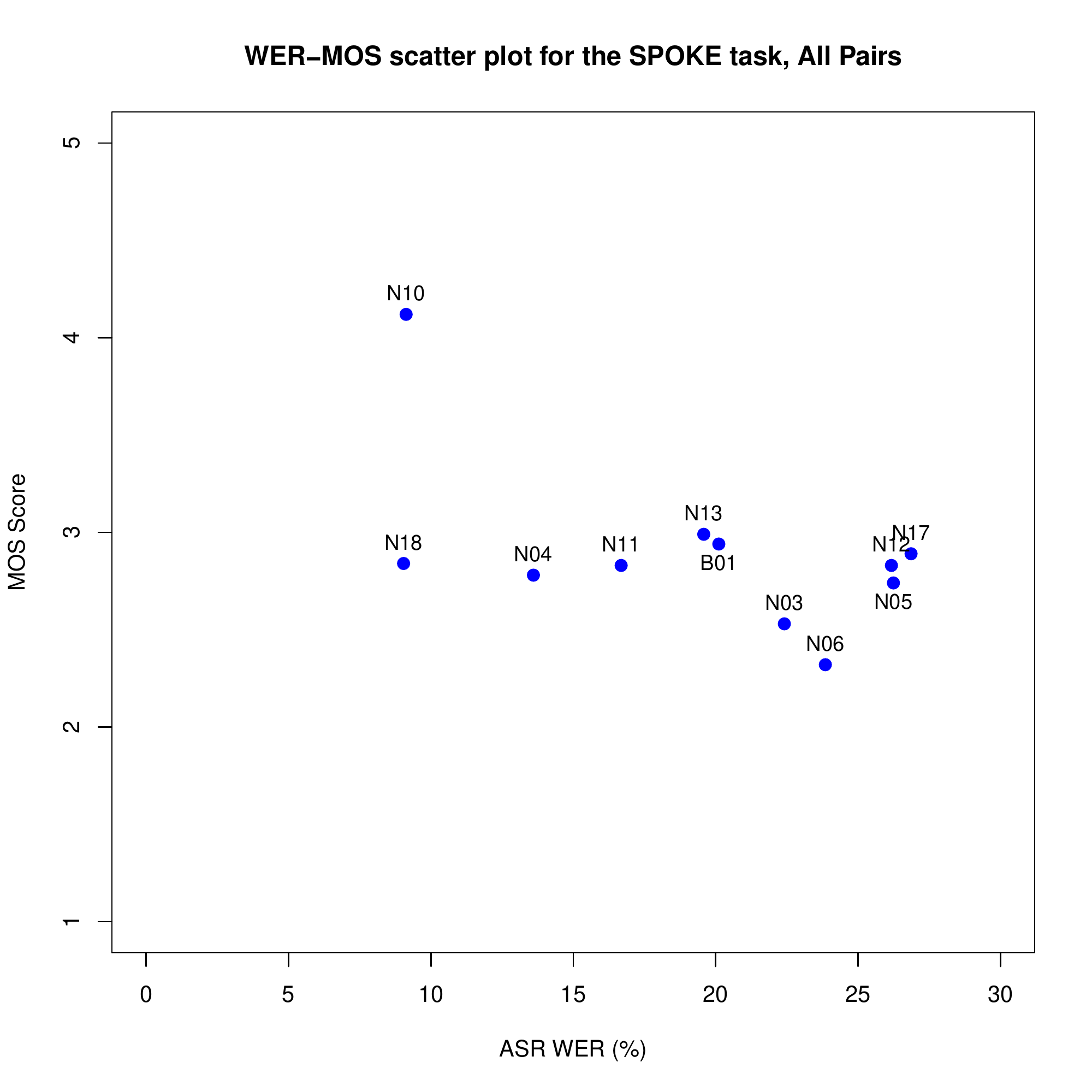}
        \vspace{-5mm}
        \caption{Scatter plot matching naturalness scores and WERs for the Spoke task  when averaging all speaker pairs.}
        \label{fig:scatter_wer_spo}
        \vspace{-5mm}
    \end{figure}

Figures~\ref{fig:scatter_wer_hub} and \ref{fig:scatter_wer_spo} show scatter plots matching naturalness scores and WERs measured by ASR for the Hub and Spoke tasks, respectively. As described earlier, we can see the reasonable correlation between the WERs of all systems and the MOS scores. This indicates that VC frameworks generally need to process linguistic information properly. 

\section{Statistical significance}

\begin{figure*}[htb]
        \centering
        \includegraphics[width=15cm]{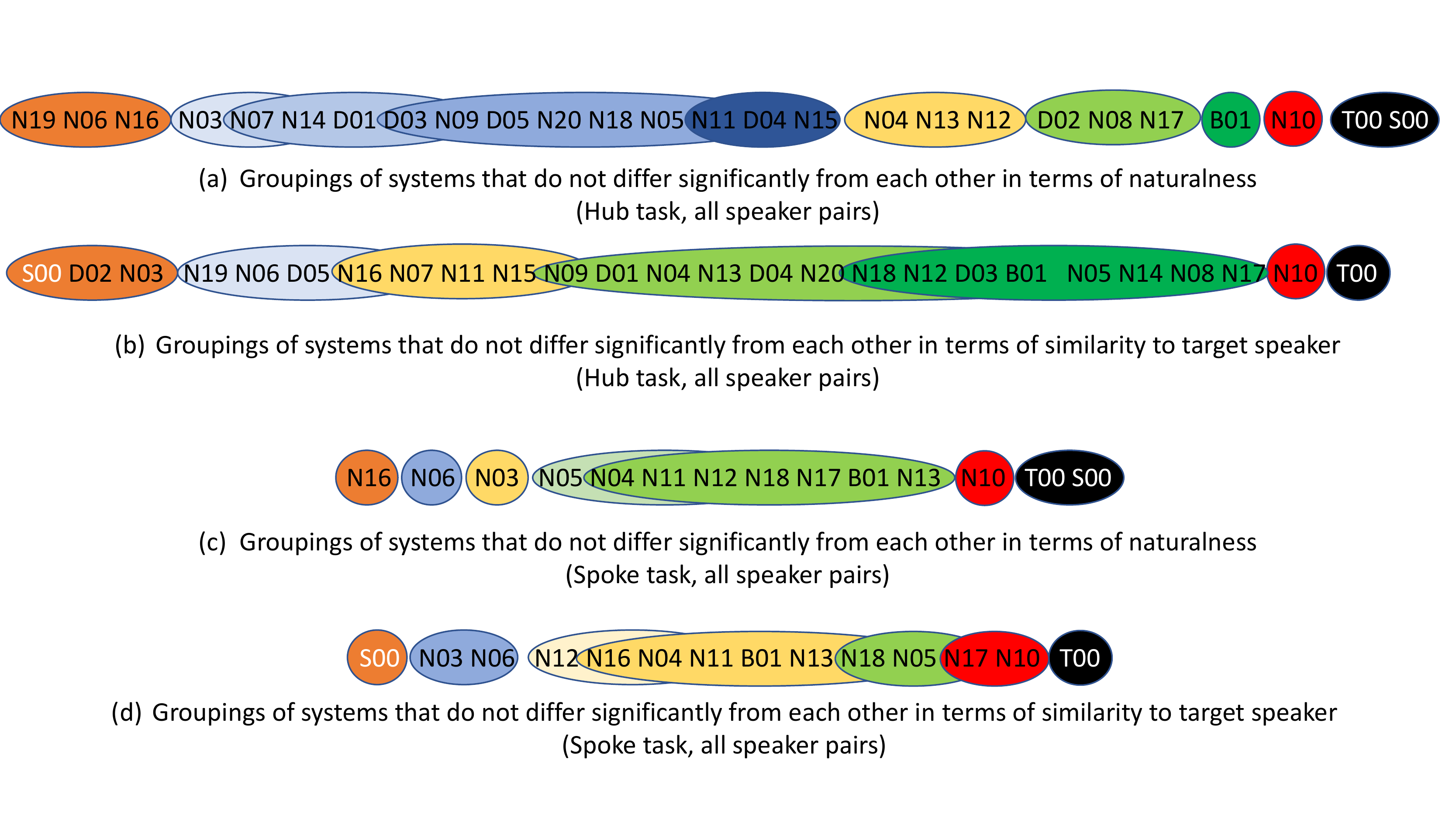}
        \caption{Significance groupings of systems}
        \label{fig:sig}
\end{figure*}

Figure \ref{fig:sig} illustrates the significance groupings of systems. Using Wilcoxon signed-rank tests with Bonferroni correction ($\alpha$ = 0.01), we performed groupings of systems that do not differ significantly from each other in terms of naturalness or similarity to target speaker. Figure \ref{fig:sig} (a) and (b) are the significance groupings for the Hub task and (c) and (d) are those for the Spoke task. 

We can see that the differences between N10 and T00, the natural speech of the target speaker, are still statistically significant. In this sense, we can say that voice conversion is not a solved problem yet. However, we can also see that N10 is statistically better than any other systems in all the evaluation apart from similarity evaluation of the Spoke task. For the Spoke task, the difference between N10 and N17 is not statistically significant in terms of similarity results of the target speaker.

\end{document}